# Soliton control in fading optical lattices

Yaroslav V. Kartashov, Victor A. Vysloukh,* and Lluis Torner

*ICFO-Institut de Ciencies Fotoniques, and Universitat Politecnica de Catalunya,*

*Mediterranean Technology Park, 08860 Castelldefels (Barcelona), Spain*

We predict new phenomena, such as soliton steering and soliton fission, in optical lattices that fade away exponentially along the propagation direction. Such lattices, featuring tunable decay rates, arise in photorefractive crystals in the wavelength range 360-400 nm. We show that the predicted phenomena offer different opportunities for soliton control.

*OCIS codes: 190.5530, 190.4360, 060.1810*

Optical solitons in periodic nonlinear media are a topic of intense investigation. Various types of solitons exist in arrays of weakly coupled waveguides [1] and optically induced lattices [2-6]. Variation of the lattice shape in the longitudinal direction opens a wealth of opportunities for soliton control [7]. Harmonic longitudinal modulation results in parametric amplification of transverse soliton swinging [7,8], and soliton dragging occurs in dynamical lattices produced by three imbalanced interfering plane waves [9,10]. Strong periodic longitudinal modulation can be used to suppress diffraction of linear beams [11], while nonlinear waveguide array built of the properly designed segments supports diffraction managed solitons [12]. In this Letter we address a new type of spatially-varying lattices that fade away along the propagation direction due to the exponential decay of the light beams that occurs in suitable wavelength bands. We show that the transverse mobility of solitons in such lattices changes with distance, a property that might find applications in all-optical switching schemes. We consider steering of single solitons and fission of soliton bound states. Also, we show how tuning the lattice decay rate allows control of output soliton position.

For the sake of generality we address beam propagation in a focusing Kerr-type medium with an imprinted transverse refractive index modulation, that is described by the nonlinear Schrödinger equation for the dimensionless field amplitude $q$:

$$i\frac{\partial q}{\partial \xi} = -\frac{1}{2}\frac{\partial^2 q}{\partial \eta^2} - q|q|^2 - pR(\eta,\xi)q. \qquad (1)$$

Here $\eta, \xi$ stand for the transverse and longitudinal coordinates scaled to the beam width and the diffraction length, respectively; parameter $p$ characterizes the lattice depth, while the function $R(\eta,\xi)$ describes the lattice profile. We consider harmonic transverse refractive index modulations and assume that the optical lattice decays exponentially with $\xi$, i.e., $R(\eta,\xi) = \cos^2(\Omega\eta)\exp(-\delta\xi)$ where $\Omega$ is the lattice frequency and $\delta$ is its decay rate (Fig. 1(c)). Such lattices can be technologically fabricated or induced optically in photorefractive materials. In the latter case one can tune the lattice parameters by changing intensities, intersection angles, and carrying wavelength of lattice-creating plane waves. For example, in SBN crystals the absorption coefficient drops off from 100 cm$^{-1}$ to 1 cm$^{-1}$ in the wavelength range 360-400 nm [13,14]. Since lattices and solitons are formed at different wavelengths, the latter do not experience absorption. The lattice decay rate can be adjusted also by changing the crystal temperature, because of the thermal shift of the absorption band edge. The transverse lattice profile does not vary despite its fading away gradually along the $\xi$ direction.

To gain intuitive insight into the soliton propagation, we start with an analytical approach by considering shallow rapidly decaying lattices. We consider evolution of sech-type beam $q_s(\eta, \xi=0) = \chi\,\text{sech}[\chi(\eta-\eta_0)]$, where $\chi$ is the form-factor and $\eta_0$ is the initial center shift. According to the inverse scattering transform, the perturbation of soliton profile $\delta q(\eta)$ results in the far-field variations of soliton form-factor $\delta\chi$ and propagation angle $\delta\alpha$ [15]:

$$\begin{aligned}\delta\chi &= \chi\int_{-\infty}^{\infty}\text{sech}[\chi(\eta-\eta_0)]\,\text{Re}(\delta q)d\eta,\\ \delta\alpha &= \chi\int_{-\infty}^{\infty}\text{sech}[\chi(\eta-\eta_0)]\tanh[\chi(\eta-\eta_0)]\,\text{Im}(\delta q)d\eta.\end{aligned} \qquad (2)$$

Rapidly decaying shallow lattice superimposes only phase modulation on the beam, so that after lattice decay soliton amplitude is given by $q_r = q_s \exp[i(p/2\delta)(1+\cos(2\Omega\eta))]$. Using expansion of exp function into series of Bessel functions, with $\mu = p/2\delta$ being a small parameter, one gets the perturbation $\delta q = q_r - q_s = [\text{J}_0(\mu)-1+2i\text{J}_1(\mu)\cos(2\Omega\eta)]q_s$,

where a phase shift independent on $\eta$ was omitted. Calculation of the first integral in Eq. (2) gives $\delta\chi = -2[1 - \mathrm{J}_0(\mu)] \approx -\mu^2/2$; thus, the form-factor only slightly diminishes. The second integral gives perturbation of propagation angle $\delta\alpha = -(4\pi\Omega^2/\chi)\mathrm{J}_1(\mu)\sin(2\Omega\eta_0)/\sinh(\pi\Omega/\chi)$. Note, that for $\mu \ll 1$ one has $\mathrm{J}_1(\mu) \approx \mu/2$, so that the far-field angle grows linearly with $p/\delta$. The sign and magnitude of $\delta\alpha$ can be tuned by the transverse shift of the input beam. Notice also that $\delta\alpha$ as a function of $\Omega$ has a single maximum, an indication that the lattice frequency could be optimized to achieve maximal soliton deflection.

To substantiate these predictions based on the above simple model, we performed direct integration of Eq. (1) with input conditions $q_{\mathrm{s}}(\eta, \xi = 0) = \chi\,\mathrm{sech}[\chi(\eta - \eta_0)]$. In order to characterize the effect of the decaying lattice on the soliton propagation path we introduce the integral soliton center $\eta_{\mathrm{int}}(\xi) = U^{-1}\int_{-\infty}^{\infty}|q|^2\eta\,d\eta$, where $U = \int_{-\infty}^{\infty}|q|^2\,d\eta$, and define its shift after $\xi_{\mathrm{end}}$ propagation units as $\delta\eta_{\mathrm{int}} = \eta_{\mathrm{int}}(\xi_{\mathrm{end}}) - \eta_{\mathrm{int}}(0)$. Fig. 1 illustrates typical soliton propagation scenarios. In the absence of an input shift ($\eta_0 = 0$) the soliton transverse position does not change with propagation. For $\eta_0 \ne 0$, the shifted soliton performs oscillations with gradually diminishing frequency inside the input channel. An estimate of oscillation frequency can be obtained from an effective particle approach (see [6] for details), where one assumes that soliton does not change its functional profile and moves like a particle inside the potential produced by the lattice, to obtain $\Omega_0(\xi) \approx [(2p\pi\Omega^2/\chi)\exp(-\delta\xi)/\sinh(\pi\Omega/\chi)]^{1/2}$. At certain distance the angle $\alpha(\xi)$ may exceed the critical value $\alpha_{\mathrm{cr}} = \Omega_0(\xi)/\Omega$ at which soliton escapes from the lattice channel (when the kinetic energy of the equivalent particle exceeds the decreasing height of the lattice potential barrier), since $\alpha_{\mathrm{cr}} \sim \exp(-\delta\xi/2)$. At this point the soliton starts moving across the lattice and it is not trapped in the neighboring lattice channels, since radiative losses as well as $\alpha_{\mathrm{cr}}$ decrease with distance. Asymptotically ($\xi \to \infty$) such beam transforms into freely walking solitons of uniform media Thus, lattice decay results in significant displacement of solitons even at short propagation distances. Depending on the lattice decay rate, solitons perform different number of oscillations and may start walking freely in diverse directions. This effect combined with tunability of optical lattice parameters (including its decay rate [13,14]) might be used for soliton steering.

Dependence of soliton center shift (further we set $\xi_{\text{end}} = 32$) on lattice frequency is shown in Fig. 2(a). In rapidly decaying lattices ($\delta \sim 2$) soliton does not oscillate but is always deflected in the same direction, dictated by the input shift $\eta_0$. There exists an optimal frequency corresponding to the largest displacement. In this regime, the analytical estimate gives accurate predictions for the dependence $\delta\eta_{\text{int}}(\Omega)$. The soliton shift goes to zero at $\Omega \to 0$ (in this limit the refractive index gradient is too small to give any appreciable acceleration to the soliton before the lattice decays) and at $\Omega \to \infty$ (in this case lattice effects are averaged out because soliton covers many lattice periods). In slowly decaying, or optically thick, lattices (e.g., $\delta = 0.25$) the soliton center performs several oscillations before soliton escapes from the input channel. Since the oscillation period depends on the lattice frequency $\Omega$, this gives rise to complex $\delta\eta_{\text{int}}(\Omega)$ dependences. Soliton displacements strongly depend on the input shift $\eta_0$ (Fig. 1(b)). The dependence $\delta\eta_{\text{int}}(\eta_0)$ is periodic with period $\pi/\Omega$, in agreement with the above thin medium approximation.

The larger the form-factor the richer the dependence $\delta\eta_{\text{int}}(\eta_0)$. This is due to the fact that the frequency of soliton oscillations inside the lattice increases with $\chi$. Thus, high-amplitude solitons launched with different shifts may perform several oscillations until the condition $\alpha = \alpha_{\text{cr}}$ is reached. In this case escape angle changes its sign several times with $\eta_0$ in contrast to the escape angle for low-amplitude solitons. This is illustrated in Fig. 2(c) that shows the dependence $\delta\eta_{\text{int}}(\chi)$. The decaying lattice does not cause substantial displacement of broad solitons with $\chi \ll 1$ covering many lattice sites, and the shift of integral center saturates for high-amplitude solitons with $\chi \gg 1$. The key property that illustrates the possibility to control the output soliton positions and escape angles is shown in Fig. 2(d). Growth of the decay rate up to $\delta \sim 1$ causes enhancement of variations of output soliton position; then the soliton center shift slowly decreases as $\delta \to \infty$. Fine tuning of the lattice decay rate results in considerable modification of output soliton position and can thus be used for soliton routing.

Decaying lattices may also be used for efficient splitting of soliton bound states. Such states form when amplitude of an input beam is $N$ times larger than that of the fundamental soliton, and they can be considered as nonlinear superposition of $N$ anti-phase solitons with form-factors $\chi_k$ ranging from 1 to $2N - 1$. The binding energy of bound states is zero in Kerr media and under action of asymmetrical perturbations they split into the fundamental solitons contained in the input beam profile. When launched into decaying lattice, bound states experience fast splitting. Importantly, in contrast to lattices invariable in $\xi$ [16], the emerging solitons move apart and effectively separate even for $p > 1$, instead of being trapped in the nearest lattice channels. In this case, by varying the decay rate of the lattice, its depth and frequency one can control the amplitudes $\chi_k$ and asymptotic escape angles $\alpha_k$. Dynamics of soliton fission in optically thick lattices is complex because single-soliton components perform several reflections inside lattice channel before separation. Here we focus on optically thin lattices, where complete splitting occurs at short distances (see Fig. 3(a), 3(b) showing fission dynamics for three-soliton bound state $q|_{\xi=0} = 3\,\text{sech}(\eta - \eta_0)$ at $\eta_0 = \pi/4\Omega$). Perturbation theory for $N$-soliton solutions (which is analogous to Eq. (2) but more tedious, see [15] for details) holds too. Figure 3(c) shows the dependence of the propagation angles on lattice frequency calculated with the aid of perturbative inverse scattering transform. Output angles $\alpha_{3,5}$ for high-amplitude solitons tend to zero at $\Omega \to 0$ and $\Omega \to \infty$. The soliton with lowest amplitude gets gradually destroyed (i.e. its form-factor $\chi_1$ tends to zero) when lattice frequency approaches $\Omega \sim 2$, so we show only a part of the curve for $\alpha_1$. There exist an optimal frequency for the largest escape angle $|\alpha_{3,5}|$. Notice that for $p/\delta < 1$ absolute values of the output angles $\alpha_k$ increase monotonically with $p/\delta$ (Fig. 3(d)). Results of direct numerical integration of Eq. (1) are in good agreement with predictions by the inverse scattering transform (compare, e.g. results of Figs. 3(a) and 3(b) with Fig. 3(c)), confirming the potential of the decaying lattices for controllable multi-soliton fission.

*Also with Universidad de las Americas – Puebla, Mexico.

# References without titles

# Figure captions

Figure 1 (color online). Propagation dynamics of soliton in decaying lattices with $\delta=0.14$ (a) and $\delta=0.2$ (b) at $p=1$, $\Omega=1$, $\chi=3$, $\eta_0=\pi/4\Omega$. (c) Profile of decaying lattice with $\delta=0.2$ and $\Omega=1$.

Figure 2. Integral soliton center shift vs (a) lattice frequency at $p=1$, $\chi=2$, $\eta_0=\pi/4\Omega$, (b) initial soliton displacement at $p=1$, $\Omega=2$, $\delta=0.5$, (c) soliton form-factor at $p=1$, $\Omega=2$, $\eta_0=\pi/4\Omega$, (d) lattice decay rate at $p=1$, $\Omega=2$, $\eta_0=\pi/4\Omega$, $\chi=2$.

Figure 3 (color online). Dynamics of decay of three-soliton bound states in lattice with $\Omega=1$ (a) and $\Omega=1.6$ (b) at $p=2$, $\delta=2$. Labels $\chi_k$ denote solitons with initial form-factors $k=1$, 3 and 5. (c) Output angles for solitons emerging after decay of bound state vs lattice frequency at $p/\delta=1$ (c) and vs $p/\delta$ at $\Omega=2$ (d).

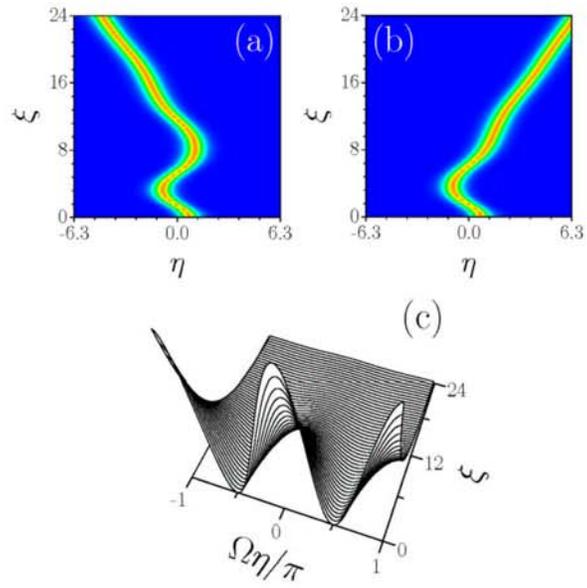

Figure 1 (color online). Propagation dynamics of soliton in decaying lattices with $\delta = 0.14$ (a) and $\delta = 0.2$ (b) at $p=1$, $\Omega=1$, $\chi=3$, $\eta_0 = \pi/4\Omega$. (c) Profile of decaying lattice with $\delta = 0.2$ and $\Omega = 1$.

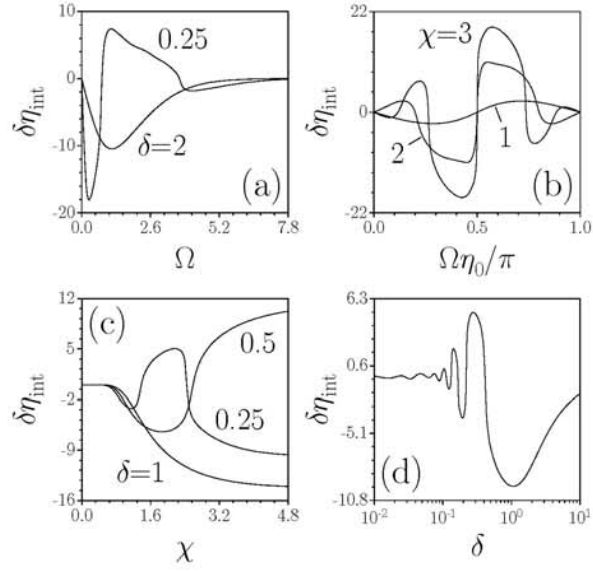

Figure 2.  Integral soliton center shift vs (a) lattice frequency at $p=1$, $\chi=2$, $\eta_0=\pi/4\Omega$, (b) initial soliton displacement at $p=1$, $\Omega=2$, $\delta=0.5$, (c) soliton form-factor at $p=1$, $\Omega=2$, $\eta_0=\pi/4\Omega$, (d) lattice decay rate at $p=1$, $\Omega=2$, $\eta_0=\pi/4\Omega$, $\chi=2$.

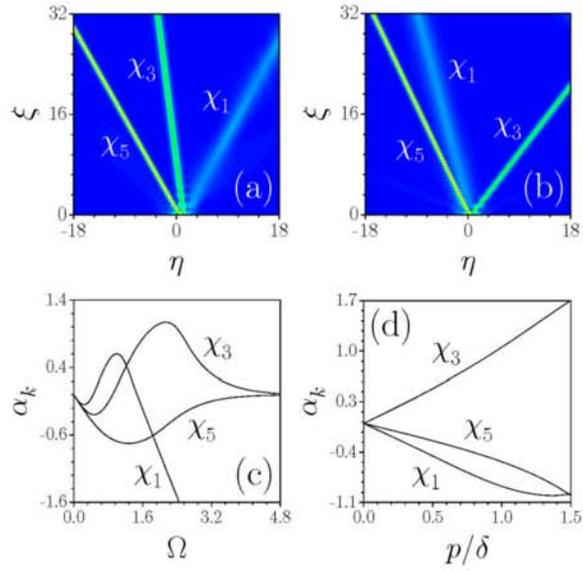

Figure 3 (color online). Dynamics of decay of three-soliton bound states in lattice with $\Omega = 1$ (a) and $\Omega = 1.6$ (b) at $p = 2$, $\delta = 2$. Labels $\chi_k$ denote solitons with initial form-factors $k = 1$, 3 and 5. (c) Output angles for solitons emerging after decay of bound state vs lattice frequency at $p/\delta = 1$ (c) and vs $p/\delta$ at $\Omega = 2$ (d).